\documentclass{cta-author}

\newtheorem{theorem}{Theorem}{}
{}
{}

\begin{document}

\supertitle{Accepted Manuscript for IET Communications}

\title{Base Station Switch-off with Mutual Repulsion in 5G Massive MIMO Networks}

\author{\au{Jiaqi Chen$^{1}$}, \au{Xiaohu Ge$^{1}$}, \au{Xueying Song$^{1}$}, \au{Yi Zhong$^{1\corr}$}}

\address{\add{1}{School of Electronic Information and Communications, Huazhong University of Science and Technology, Wuhan 430074, Hubei, P. R. China.}
\email{yzhong@hust.edu.cn}}

\begin{abstract}
When small cells are densely deployed in the fifth generation (5G) cellular networks, switching off a part of base stations (BSs) is a practical approach for saving energy consumption considering the variation of traffic load. The small cell network with the massive multi-input multi-output (massive MIMO) system is analyzed in this paper due to the dense deployment and low power consumption. Based on the BS switch-off strategy with distance constraints, the energy and coverage efficiency are investigated to illustrate the performance of the BS switch-off strategy. Simulation results indicate that the energy efficiency and coverage efficiency of the proposed strategy are better than the random strategy. The energy efficiency increases with the BS intensity and the minimal distance, and a maximum coverage efficiency can be achieved with the increase of the BS intensity and the minimum distance. In this case, the optimal BS switch-off strategy can be designed under this work in the actual scene.
\end{abstract}

\maketitle

\section{Introduction}\label{sec1}

With the development of the fifth generation (5G) mobile communication system, key technologies such as the massive multi-input multi-output (massive MIMO) technology, millimeter wave technology and small cell networks, are emerging as reasonable solutions for 5G cellular networks~\cite{Ge16M}. To deploy the large antenna array of the massive MIMO system in a base station (BS), the antenna spacing needs to be reduced. The millimeter wave technology matches the massive MIMO system very well due to the short wavelength of the millimeter wave. Considering that the millimeter wave signals are sensitive to the blockages in the propagation environment, the propagation distance of a BS with the massive MIMO system is reduced compared with the traditional microwave BS. In this case, a large number of low power BSs have to be deployed densely in 5G cellular networks to meet the requirement of seamless coverage~\cite{Ge17M}.

Moreover, the energy efficiency needs to be optimized to satisfy the requirement of the future 5G green networks~\cite{Wang16M}. BS switch-off strategies are practical approaches to save energy in green cellular networks since 80\% energy of cellular networks is consumed at macro BSs~\cite{Ashraf11M,Celebi17C}. Considering that the cooling unit in a small cell BS is abolished, the primary power consumption of the small cell BS is used for transmitting signals and computing~\cite{Ge17M}. The power consumption model of a small cell BS is different from a macro BS. In this case, it is essential to analyze the impacts of the BS switch-off strategies for small cell networks with the massive MIMO system.

Various studies have been carried out for BS switch-off strategies for greener 5G small cell networks~\cite{Soh13C,Li16J,Li17J,Kashef16J,Le-The17L,Cai16J}. Regarding the control conditions for the selection of the switched BSs, the BS switch-off strategies can be categorized into random strategy, traffic-aware strategy, and distance-aware strategy. Two sleeping mode strategies, {\em i.e.}, random and strategic sleeping modes, were proposed and analyzed for the energy efficiency of cellular networks in~\cite{Soh13C}, where the probabilities that a BS continues to work in the two sleeping modes are configured as a constant and a function of the traffic. In~\cite{Li16J}, a BS was considered to be switched when there is no user equipment (UE) in the cell, such that the probability that a BS is switched is a function of the UE density. The above three strategies can be regarded as the random strategies due to the independent decision making from the actual network deployment. The traffic load is an important factor for BS switch-off strategies~\cite{Zhong17J,Zhong17M}. A concept called traffic-aware cell management was innovated in~\cite{Li17J}, which involves cell division, cell death, and cell migration to represent adaptations of networks, where the state transitions of BSs are controlled. The direction of arrival (DOA) was adopted in~\cite{Li17J} for distributed decision making. A balanced dynamic planning approach that accounts for quality of service (QoS) requirements both in the uplink and downlink was proposed in~\cite{Kashef16J} to specify the BS switching decision, which was formulated as a two-timescale average reward Markov decision process with finite horizon. The traffic-aware cell management and balanced dynamic planning approach are traffic-aware strategies because the decision making of these strategies are determined by the users' traffic requirements and QoS requirements. Given that the best signal-to-interference ratio (SIR) distribution can be achieved when BSs are located on a hexagonal layout, the authors in \cite{Le-The17L} studied applying cell switch-off algorithms to irregular network layouts with the objective of making the active BS locations as regular as possible, regardless of the irregularity of the original network layout. An optimal location-based operation scheme was proposed in~\cite{Cai16J} by gradually switching off the small cell BSs closer to the macro BS, in order to reduce the total power consumption of the heterogeneous networks (HetNets) while keeping the marco BS on to avoid any service failure outside active small-cells. The decision making of the optimal location-based operation scheme is associated with the distance between the candidate small cell BSs and the macro BS, which is regarded as a distance-aware strategy. However, the impact of the relationship between switched BSs and active BSs on the performance of the thinning network could not come into sharp focus in all above strategies.

The BS switch-off strategy has been emerged as a viable solution to enhance the overall network energy efficiency by inactivating the unutilized BSs. However, it affects the performance of UEs associated with the switched BSs depending on the BS association scheme, channel conditions toward the active BSs, and traffic loads at the active BSs. In this case, the network energy efficiency and coverage probability of the BS switch-off strategy are the main parameters to be analyzed and optimized~\cite{Tabassum14J,Ebrahim17J,Yu16J,Bousia16J,Zhang15J,Renga18J}. The impact of different multi-user scheduling schemes was analyzed in~\cite{Tabassum14J} on the channel access probability and spectral efficiency of the sleeping cell users given a certain BS-switching-off pattern, and numerical results demonstrated the efficacy of the maximum mean channel access probability-based association scheme in non-uniform traffic load scenarios. The capacity and energy consumption metrics of small-cell networks that are enabled with sleep mode functionality were investigated in~\cite{Ebrahim17J} to systematically and accurately identify the potential sleep mode cells that can maximize the spectrum reuse efficiency without the need for an exhaustive search. By considering the switching cost, an energy saving problem of BSs in cellular networks was formulated as the minimum energy cost problem and solved in two steps in~\cite{Yu16J}. In~\cite{Bousia16J} by employing auction tools and novel bidding strategies, a switching off scheme was introduced that allows mobile networks to reduce expenditures, and was evaluated in terms of energy efficiency and cost metrics for various conditions (different bidding levels and behaviors). In~\cite{Zhang15J} the expected sleeping ratio of small cells with the time-varying traffic was obtained for two small cell sleeping schemes (random and repulsive schemes), under which the UEs of the sleeping small cells are vertically offloaded to the macro BS-layer for outage probability guarantee. The proposed energy management policy with dynamically switching off unneeded BSs in \cite{Renga18J}, applied in a renewable energy-powered mobile access network to respond to the smart gird requests was highly effective in reducing the operational cost, that may decrease by up to more than 100\% under proper setting of operational parameters.

In our previous work~\cite{Jia16C}, a BS switch-off strategy with distance constraints was proposed for reducing the energy consumption in the point to point (P2P) cellular networks. The mutual repulsion among active BSs in the proposed switch-off strategy was analyzed and influences the performance of cellular networks which cannot be ignored. In \cite{Du17J}, we proposed a multi-user multi-antenna cellular network model with the aforementioned minimum distance constraint for adjacent base stations, where the locations of all BSs were modeled by the hard-core point process (HCPP). Above studies of BS switch-off strategies are based on the traditional cellular networks. However, small cell networks with the massive MIMO technology is emerging as the future 5G network architecture. Considering that the cooling unit in a small cell BS is abolished, the primary power consumption of the small cell BS is used for transmitting signals and computing. The energy saving caused by switching off a small cell BS is reduced compared with the scenario with macro BSs. Moreover, the adoption of massive MIMO technology at the small cell BS can provide the higher capability to serve more users compared with the MIMO or P2P BS, which results in the assurance of the users¡¯ QoS. Therefore, it is essential to analyze the impacts of the BS switch-off strategies for small cell networks with the massive MIMO system. And the trade-off between the energy saving and coverage loss in small cell networks with the massive MIMO system need to be reassessed.

In this paper, we introduce the hard core point process (HCPP) to model the BS distribution of the thinning small cell network when the BS switch-off strategy with mutual repulsion is adopted. The minimum distance between active BSs is an important factor for the BS switch-off strategy in 5G small cell networks. By formulating the energy efficiency and coverage efficiency of the thinning small cell network with the massive MIMO system, the limit of coverage efficiency is obtained in 5G green small cell networks with massive MIMO systems. The contributions of this paper are summarized as follows:
\begin{enumerate}[(iii)]
  \item[(i)] The HCPP model is introduced to model the thinning small cell network which adopts the mutual repulsive BS switch-off strategy. Moreover, the small cell network with massive MIMO system is considered.
  \item[(ii)] The energy and coverage efficiency of the small cell network with massive MIMO system are derived to investigate the performance of the proposed BS switch-off strategy.
  \item[(iii)] Considering the BS switch-off strategy with distance constraints, simulation results indicate that the limit for the coverage efficiency is existing.
\end{enumerate}

The remainder of this paper is organized as follows. Section II introduces the system model. The interference model of the small cell network is presented in Section III. The energy efficiency of the thinning small cell network is proposed in Section IV. The coverage efficiency of the thinning small cell network is analyzed in Section V. Simulation results are shown in Section VI. Finally, conclusions are drawn in Section VII.

\section{System Model}\label{sec2}

To save the energy consumption of small cell networks with light traffic load, the BSs switch-off strategy is one of main approaches to improve the energy efficiency of small cell networks. In this case, a part of adjacent small cell BSs are selected to switch off to guarantee the coverage of small cell networks. Therefore, a BS switch-off strategy with mutual repulsion and the channel model of the massive MIMO system are introduced respectively in the following.

\subsection{BS Switch-off Strategy with Mutual Repulsion}
A small cell network with the massive MIMO technology is considered in this paper. We assume that all BSs are randomly located in the infinite plane $\mathbb{R}^2$. The locations of all BSs are assumed to be governed by a homogeneous Poisson point process (PPP) with intensity $\lambda_B$, {\em i.e.}, $\Pi_{\text{PPP}}=\{x_i,\;i=1,2,\ldots\}$, where $x_i$ denotes the two dimensional Cartesian coordinate of the $i$--th BS ${\text{BS}}_i$. Considering that studies have shown that there are high fluctuations in traffic demand over space and time in cellular networks~\cite{Auer11M}, {\em e.g.}, the traffic demands in urban and rural areas or traffic demands in day and night time are entirely different, there is potential in energy savings by adapting the BS switch-off strategy to the demanded traffic.

\begin{figure}
  \centering\includegraphics[width=8cm]{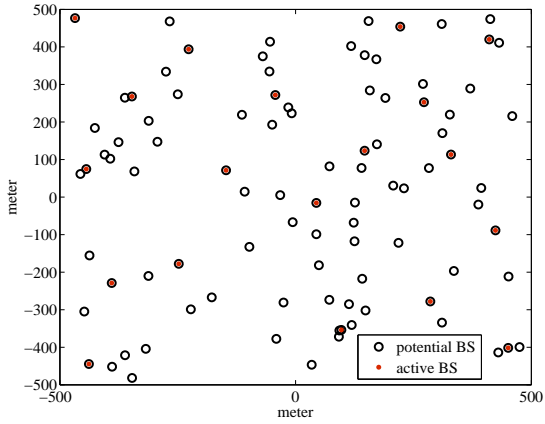}\\
  \caption{The BS distribution in a small cell network with the mutually repulsive BS switch-off strategy. The hollow circles are all the BSs in the small cell network. The red solid dots are the active BSs after adopting the mutually repulsive BS switch-off strategy.}\label{f1}
\end{figure}

In this paper, a mutually repulsive BS switch-off strategy is proposed based on the distance constraint between any two BSs. When the traffic demand of the small cell network is heavy, all BSs are switched on to work. When the traffic demand of the small cell network is light, a part of BSs should be switched off to save energy. In this case, all BSs calculate the current traffic load to peak traffic load ratios denoted as $\Phi\left(x_i\right)\left(\;i=1,2,\ldots\right)$, which are assumed to be independently uniformly distributed in the range of $\left[0,1\right]$. The BS switch-off process in the small cell network is regarded as a dependent thinning process of the PPP with a minimum distance $\delta$ between active BSs. The dependent thinning retains the active BS ${\text{BS}}_i$ of PPP with mark $\Phi\left(x_i\right)$ if the disk $d\left(x_i,\delta\right)$ contains no BSs with marks larger than $\Phi\left(x_i\right)$, where $d\left(x_i,\delta\right)$ is a disk region with central point $x_i$ and the radius $\delta$~\cite{Stoyan95B}. After adopting the BS switch-off strategy, the active BSs of the small cell network can be governed by a Mat\'{e}rn hard-core process of Type II~\cite{ElSawy12C} in this paper, which represents a special case of HCPP. Moreover, the thinned process $\Pi_{\text{HCPP}}$, {\em i.e.}, the Mat\'{e}rn hard-core process~\cite{Stoyan95B} is defined by
\begin{equation}\label{hcpp}
  \Pi_{\mathrm{HCPP}}\!=\!\left\{x \in \Pi_{\text{PPP}}: \!\Phi\left(x\right) > \Phi\left(x^{\ast}\right),\;\forall x^{\ast}\!\in\! \Pi_{\text{PPP}} \cap d\left(x,\delta\right)\right\}.
\end{equation}
The distributions of BSs with and without the BS switch-off strategy in a small cell network are illustrated in Fig.~\ref{f1}. Assumed that each small cell includes $L$ UEs in the original small cell network, the number of the UEs in a cell is regarded as $K=\frac{L\lambda_B}{\lambda^{\ast}}$ when adopting the mutually repulsive BS switch-off strategy, where $\lambda^{\ast}$ denotes the density of the active BSs in the thinning small cell as well as the intensity of the HCPP.

\subsection{Channel Model}
In this paper, we assume that each small cell in the thinning small cell network includes one BS equipped with $M$ antennas and $K$ UEs equipped with a single antenna. Considering that the massive MIMO technology is adopted at BSs, the number of antennas at the BS is obviously larger than the number of UEs. In this case, the channel model~\cite{Ngo13J} between the desired BS and UEs is
\begin{equation}\label{1}
\mathbf{H}=\mathbf{G}\mathbf{D}^{\frac{1}{2}},
\end{equation}
where $\mathbf{G}$ is a $M \times K$ matrix denoting the fast fading coefficients between a BS and $K$ UEs, the elements of which are independent identical distribution ({\em i.i.d.}) complex Gaussian random variables with unit variance and zero mean. The element $[\mathbf{G}]_{mk} = g_{mk}\;(m=1,2,\ldots,M;k=1,2,\ldots,K)$ located at the $m$--th row and the $k$--th column of matrix $\mathbf{G}$ is the fast fading coefficient between the $m$--th antenna at the desired BS and the $k$--th UE. $\mathbf{D}$ is a $K \times K$ diagonal matrix, and the diagonal element, {\em i.e.}, $[\mathbf{D}]_{kk} = \beta_k$, is the large scale fading coefficient between the desired BS and the $k$--th UE. Moreover, the large scale fading coefficient is extended by $\beta_k=\frac{\omega_k}{r_k^{\alpha}}$, where $\omega_k$ is the shadowing effect in wireless channels and is calculated by $\omega_k=e^{\frac{s}{10}}$, the random variable $s$ is governed by the Gaussian distribution with zero mean and variance $\sigma_s^2$, {\em i.e.}, $s\sim N(0, \sigma_s^2)$, $r_k$ is the distance between the desired BS and the $k$--th UE, $\alpha$ is the path loss coefficient.

The received signal vector $\mathbf{y}_i$ of $K$ UEs in the $i$--th small cell is expressed as
\begin{align}
& \mathbf{y}_i=\mathbf{H}_{ii}^{\text{T}}\mathbf{x}_{i}+\sum_{u\neq i}\mathbf{H}_{ui}^{\text{T}}\mathbf{x}_{u}+\mathbf{n}, \label{yi} \\
& \mathbf{x}_{i}=\sqrt{P_f}\mathbf{F}_{i}\mathbf{s}_{i}, \label{xi}\\
& \mathbf{x}_{u}=\sqrt{P_f}\mathbf{F}_{u}\mathbf{s}_{u}, \label{xu}
\end{align}
where $\mathbf{H}_{ii}^{\text{T}}$ is the channel matrix between the desired BS $\text{BS}_i$ and the $K$ UEs in the $i$--th small cell with the transposition operation $(\cdot)^{\text{T}}$ at the matrix, $\mathbf{x}_{i}$ is the $M \times 1$ signal vector transmitted by $\text{BS}_i$, $\mathbf{H}_{ui}^{\text{T}}$ is the channel matrix between the interfering BS $\text{BS}_u$ and the $K$ UEs in the $i$--th small cell, $\mathbf{x}_{u}$ is the signal vector transmitted by the interfering $\text{BS}_u$, $\mathbf{n}$ is the $K \times 1$ noise vector at the UEs in the $i$--th small cell and the noise powers at all UEs are the same and equal to $\sigma_n^2$. $\mathbf{F}_{i}$ and $\mathbf{F}_{u}$ are the $M \times K$ precoding matrixes at $\text{BS}_i$ and $\text{BS}_u$, respectively. $\sqrt{P_f}\mathbf{s}_{i}$ and $\sqrt{P_f}\mathbf{s}_{u}$ are the $K \times 1$ original signal vectors without precoding at BSs, where $P_f$ is the signal power, $\mathbf{s}_{i} \sim \mathcal{CN}(0, 1)$ and $\mathbf{s}_{u} \sim \mathcal{CN}(0, 1)$ are the independent data streams at $\text{BS}_i$ and $\text{BS}_u$, respectively.

Assume that the pilot signals in different small cells are transmitted at the same frequency and the pilot signals of $K$ UEs are orthonormal each other in a small cell. In this case, the inter-cell pilot pollution is only considered and the intra-cell pilot pollution can be ignored. Then, the estimated channel matrix between the desired BS and the $K$ UEs in the $i$--th small cell is
\begin{equation}\label{hatHii}
  \widehat{\mathbf{H}}_{ii}=\sqrt{P_p}\left(\mathbf{H}_{ii}+\sum_{u\neq i}\mathbf{H}_{iu}\right),
\end{equation}
where $\mathbf{H}_{iu}$ denotes the channel matrix between the desired BS $\text{BS}_i$ and the $K$ UEs in the $u$--th small cell, $P_p$ is the pilot signal power.

When the matched filter precoding scheme~\cite{Lee12C} is adopted at BSs, the precoding matrices $\mathbf{F}_{i}$ and $\mathbf{F}_{u}$ used at the desired BS and the interfering BS are given by
\begin{align}
& \mathbf{F}_{i}=\widehat{\mathbf{H}}_{ii}^{\ast}=\sqrt{P_p}\left(\mathbf{H}_{ii}+\sum_{u\neq i}\mathbf{H}_{iu}\right)^{\ast}, \label{Fi} \\
& \mathbf{F}_{u}=\widehat{\mathbf{H}}_{uu}^{\ast}=\sqrt{P_p}\left(\mathbf{H}_{uu}+\sum_{i\neq u}\mathbf{H}_{ui}\right)^{\ast}, \label{Fu}
\end{align}
where $(\cdot)^{\ast}$ is the conjugation operation at the matrix. Substituting (\ref{Fi}) and (\ref{Fu}) into (\ref{xi}) and (\ref{xu}), the received signal vector $\mathbf{y}_i$ of $K$ UEs in the $i$--th small cell is derived as
\begin{align}\label{yi2}
  \nonumber &\mathbf{y}_i =  \sqrt{P_{f}P_p}\mathbf{H}_{ii}^{\text{T}}\mathbf{H}_{ii}^{\ast}\mathbf{s}_{i}+\sqrt{P_{f}P_p}\sum_{u^{'}\neq i}\mathbf{H}_{ii}^{\text{T}}\mathbf{H}_{iu^{'}}^{\ast}\mathbf{s}_{i}\\
 &+\!\sqrt{P_{f}P_p}\sum_{u\neq i}\mathbf{H}_{ui}^{\text{T}}\mathbf{H}_{uu}^{\ast}\mathbf{s}_{u}\!+\! \sqrt{P_{f}P_p}\sum_{u\neq i}\sum_{i^{'}\neq u}\mathbf{H}_{ui}^{\text{T}}\mathbf{H}_{ui^{'}}^{\ast}\mathbf{s}_{u}\!+\!\mathbf{n},
\end{align}
where $(\cdot)^{\dagger}$ is the conjugation transposition operation at the matrix. When $M$ approaches infinite and $M \gg K$, the following results are derived by~\cite{Marzetta10J}
\begin{equation}
  \frac{1}{M}\mathbf{H}_{ii}^{\text{T}}\mathbf{H}_{ii}^{\ast}=\frac{1}{M}\mathbf{H}_{ii}^{\text{T}}\left(\mathbf{H}_{ii}^{\text{T}}\right)^{\dagger} \rightarrow \mathbf{D}_{ii},
\end{equation}
\begin{equation}
  \frac{1}{M}\sum_{i^{'} \neq u}\mathbf{H}_{ui}^{\text{T}}\mathbf{H}_{ui^{'}}^{\ast}=\frac{1}{M}\sum_{i^{'} \neq u}\mathbf{H}_{ui}^{\text{T}}\left(\mathbf{H}_{ui^{'}}^{\text{T}}\right)^{\dagger} \rightarrow \mathbf{D}_{ui}.
\end{equation}

When massive MIMO antennas are assumed to be equipped at BSs, the fast fading effect in wireless channels can be ignored based on the results in~\cite{Marzetta10J}. As a consequence, (\ref{yi2}) can be simplified as
\begin{equation}\label{yi3}
  \mathbf{y}_i \!\rightarrow \!M\sqrt{P_{f}P_p}\mathbf{D}_{ii}\mathbf{s}_{i}\!+\!M\sqrt{P_{f}P_p}\sum_{u \neq i}\mathbf{D}_{ui}\mathbf{s}_{u}\!+\!\mathbf{n},\;\text{as}\;M\!\rightarrow\!\infty,
\end{equation}
where $\mathbf{D}_{ii}$ and $\mathbf{D}_{ui}$ are $K \times K$ diagonal matrices and the diagonal elements of $\mathbf{D}_{ii}$ and $\mathbf{D}_{ui}$ are denoted by $[\mathbf{D}_{ii}]_{kk}=\beta_{iki}$ and $[\mathbf{D}_{ui}]_{kk}=\beta_{uki}$, which are the large scale fading coefficients over wireless channels. Moreover, the large scale fading coefficients are calculated by $\beta_{uki}=\frac{\omega_{uki}}{r_{uki}^{\alpha}}$ and $\beta_{iki}=\frac{\omega_{iki}}{r_{iki}^{\alpha}}$, where $\omega_{uki}$ and $\omega_{iki}$ are shadowing effects over wireless channels, $r_{uki}$ is the distance between the interfering BS ${\text{BS}}_u$ and the $k$--th UE in the $i$--th small cell, and $r_{iki}$ is the distance between the desired BS ${\text{BS}}_i$ and the $k$--th UE in the $i$--th small cell, $\alpha$ is the path loss coefficient over wireless channels. Furthermore, the received signal at the $k$--th UE in the $i$--th small cell is
\begin{equation}\label{yik}
  y_{ik}=M\sqrt{P_{f}P_p}\beta_{iki}s_{ik}+M\sqrt{P_{f}P_p}\sum_{u \neq i}\beta_{uki}s_{uk}+n_k,
\end{equation}
where $s_{uk}$ is the $k$--th element in the vector $\mathbf{s}_{u}$ and is the independent data stream transmitted
for the $k$--th UE in the $u$--th small cell, $n_k$ is the noise of the $k$--th UE. Hence, the interference power received at the $k$--th UE in the $i$--th small cell is
\begin{align}\label{Iik}
  \nonumber I_{ik} &= \left(M\sqrt{P_{f}P_p}\sum_{u \neq i}\beta_{uki}s_{uk}\right)^{\dagger}\left(M\sqrt{P_{f}P_p}\sum_{u \neq i}\beta_{uki}s_{uk}\right) \\
  \nonumber &= M^{2}P_fP_p\sum_{u \neq i}\beta_{uki}^{2}s_{uk}^{\dagger}s_{uk} \\
  &= M^{2}P_fP_p\sum_{u \neq i}\beta_{uki}^{2}.
\end{align}

\section{Interference Model}\label{sec3}

When the BS switch-off strategy is adopted in the small cell network, the active BSs are included by a set $\Pi_{\mathrm{HCPP}}$. Then the interference power received at the $k$--th UE in the $i$--th small cell is expressed as
\begin{equation}\label{Iik2}
  I_{ik}=M^{2}P_fP_p\sum\limits_{u \neq i}\frac{\omega^2_{uki}}{r_{uki}^{2\alpha}},
\end{equation}
where the distance between the interfering BS ${\mathrm{BS}}_u$ and the $k$--th UE ${\mathrm{UE}}_{ik}$ in the $i$--th cell is denoted as $r_{uki} =\left|x_{{\mathrm {BS}}_u}-x_{{\mathrm {BS}}_i}-x_{{\mathrm {int}}}\right|$, $x_{{\mathrm {BS}}_u}$ and $x_{{\mathrm {BS}}_i}$ respectively denote the locations of the ${\mathrm {BS}}_u$ and ${\mathrm {BS}}_i$, and $x_{{\mathrm int}}$ denotes the distance vector form the desired BS ${\mathrm {BS}}_i$ to the ${\mathrm{UE}}_{ik}$. The average interference of an UE in the small cell network can be expressed as
\begin{equation}\label{eIk}
  I_{k\_{\rm avg}}=\lim\limits_{\mathbb S \rightarrow\mathbb R^2}\frac{\mathbb{E}\left[\sum\limits_{x_i\in\mathbb S}{I_{ik}}\right]}{\int_{\mathbb S}\zeta^{(1)}\mathrm{d}x},
\end{equation}
where $\mathbb{E}\left[\cdot\right]$ is the expectation operation, $\mathbb S$ is a finite region in the infinite plane $\mathbb S \subseteq \mathbb R^2$, and $\zeta^{(1)}$ denotes the probability that a BS is active when the BS switch-off strategy is adopted. The probability that there is a BS in the infinitesimal small region $\mathrm{d}x$ is calculated as $\zeta^{(1)}\mathrm{d}x$ for the HCPP. The denominator part of (\ref{eIk}) denotes the average number of active BSs in the small cell network.

For the PPP model, the number of points in a circle with the radius $\delta$ is a Poisson random variable with mean $\lambda_B\pi\delta^2$. Based on the results in~\cite{Du17J}, the first moment of HCPP is
\begin{equation}\label{s1}
  \zeta^{(1)}=\lambda_B\int^{1}_{0}e^{-\lambda_B\pi\delta^2t}\mathrm{d}t=\frac{1-e^{-\lambda_B\pi
  \delta^2}}{\pi\delta^2},
\end{equation}
which is the probability that a BS is active. Moreover, the second moment of HCPP which denotes the probability that two points at different locations are retained, is expressed as
\begin{subequations}
\begin{align}\label{s2}
&\zeta^{(2)}(r) = \lambda_B^2\varphi(r), \\
&\varphi(r) \!= \!\left\{\!
                   \begin{array}{ll}
                     \frac{2V_\delta(r)\left(1-e^{-\lambda_B\pi\delta^2}\right)-2\pi\delta^2\left(
                     1-e^{-\lambda_BV_\delta(r)}\right)}{\lambda_B^2\pi\delta^2V_\delta(r)(V_\delta
                     (r)-\pi\delta^2)}, & \hbox{$r\!>\!\delta$} \\
                     0, & \hbox{$r\!\leq\!\delta$}
                   \end{array},
                 \right.
  \\
&V_\delta(r) \!=\! \left\{\!
                    \begin{array}{ll}
                      2\pi\delta^2\!-\!2\delta^2\arccos\left(\frac{r}{2\delta}\right)\!+\!r\sqrt{\delta^2-
                      \frac{r^2}{4}}, & \hbox{$0\!<\!r\!<\!2\delta$} \\
                      2\pi\delta^2, & \hbox{$r\!\geq \!2\delta$}
                    \end{array},
                  \right.
\end{align}
\end{subequations}
where $r$ is the distance between two points which are located in the infinitesimally small regions, {\em i.e.}, $\mathrm d{x_1}$ and $\mathrm d{x_2}$. Based on (\ref{s2}), the probability that the two infinitesimally small regions $\mathrm d{x_1}$ and $\mathrm d{x_2}$ exist the active BSs is $\zeta^{(2)}(r)\mathrm d{x_1}\mathrm d{x_2}$, when the distance between $\mathrm d{x_1}$ and $\mathrm d{x_2}$ is $r$. In this case, the expectation of the total interference power of the thinning small cell network in the area $\mathbb S$ is derived as
\begin{align}\label{eIk2}
\nonumber  &\mathbb{E}\left[\sum\limits_{x_i\in\mathbb S}{I_{ik}}\right]\\
  &\!=\!\int\limits_{\mathbb S}\!\int\limits_{\mathbb S}\!M^{2}P_fP_p\frac{\mathbb{E}_{\omega_{uki}}\left[\omega^2_{uki}\right]}{\left|x_1\!\!-\!\!x_2\!\!-\!\!x_{{\mathrm {int}}}\right|^{2\alpha}}\zeta^{(2)}\left(\left|x_1\!\!-\!\!x_2\!\!-\!\!x_{{\mathrm {int}}}\right|\right)\mathrm d{x_1}\mathrm d{x_2}.
\end{align}
As a result, the average interference of the $k$--th UE $\mathrm {UE}_{ik}$ in the $i$--th cell of the thinning small cell network can be expressed as
\begin{align}\label{eI}
 \nonumber &I_{k\_{\rm avg}} \\
 \nonumber&= \lim\limits_{\mathbb S \rightarrow\mathbb R^2}\frac{\int\limits_{\mathbb S}\int\limits_{\mathbb S}\frac{M^{2}P_fP_p\mathbb{E}_{\omega_{uki}}\left[\omega^2_{uki}\right]}{\left|x_1\!-\!x_2\!-\!x_{{\mathrm {int}}}\right|^{2\alpha}}\zeta^{(2)}\left(\left|x_1\!-\!x_2\!-\!x_{{\mathrm {int}}}\right|\right)\mathrm d{x_1}\mathrm d{x_2}}{\int_{\mathbb S}\zeta^{(1)}\mathrm{d}x}\\
\nonumber &= \lim\limits_{\mathbb S \rightarrow\mathbb R^2}\frac{\left|\mathbb S\right|\int\limits_{\mathbb R^2}M^{2}P_fP_p\frac{\mathbb{E}_{\omega_{iu}}\left[\omega^2_{uki}\right]}{\left|-x_2-x_{{\mathrm {int}}}\right|^{2\alpha}}\zeta^{(2)}\left(\left|-x_2-x_{{\mathrm {int}}}\right|\right)\mathrm d{x_2}}{\zeta^{(1)}\left|\mathbb S\right|}\\
&=\frac{M^{2}P_fP_p\mathbb{E}_{\omega_{iu}}\left[\omega^2_{uki}\right]}{\zeta^{(1)}}\int\limits_{\mathbb R^2}\frac{\zeta^{(2)}\left(\left|x\right|\right)}{\left|x+x_{{\mathrm {int}}}\right|^{2\alpha}}\mathrm d{x},
\end{align}
where $\left|\mathbb S\right|$ denotes the area of the region $\mathbb S$.

\section{Energy Efficiency}\label{sec4}

Based on (\ref{yik}), the desired received signal power of $\mathrm{UE}_k$ in the $i$--th small cell is expressed as
\begin{align}\label{pk}
\nonumber  P_{ik} & = \left(M\sqrt{P_{f}P_p}\beta_{iki}s_{ik}\right)^{\dag}\left(M\sqrt{P_{f}P_p}\beta_{iki}s_{ik}\right)\\
   & = M^2P_fP_p\beta_{iki}^2.
\end{align}
Then the achievable rate of $\mathrm{UE}_k$ is derived as
\begin{equation}\label{rk}
  R_{ik}  = \mathrm{log}_2\left(1+\frac{P_{ik}}{I_{ik}+\sigma_n^2}\right)= \mathrm{log}_2\left(1+\frac{M^2P_fP_p\beta_{iki}^2}{I_{ik}+\sigma_n^2}\right).
\end{equation}

\begin{theorem}\label{T1}
The lower bound for the average achievable rate of $\mathrm{UE}_k$ in the $i$--th small cell with the HCPP model can be given as
\begin{equation}\label{er2}
  \begin{split}
    &R_{ik\_{\rm avg}}\left(x_{{\mathrm {int}}}\right)\!=  \\
    & \int\limits_0^{\infty}\!\mathrm{log}_2\!\left(\!1\!\!+\!\!\frac{M^{2}P_fP_p\omega^2\zeta^{(1)}e^{-2\sigma_s^2}
    \left|x_{\mathrm{int}}\right|^{-2\alpha}}{\left(M^{2}P_fP_p\!\int\limits_{\mathbb R^2}\frac{\zeta^{(2)}\left(\left|x\right|\right)}{\left|x\!+\!x_{{\mathrm {int}}}\right|^{2\alpha}}\mathrm d{x}\!+\!\zeta^{(1)}\!e^{-2\sigma_s^2}\sigma_n^2\right)}\right)\\
&\times\frac{5\mathrm {exp}\left(-\frac{25\ln^2(\omega)}{2\sigma_s^2}\right)}{\omega\sqrt{2\pi}\sigma_s}\mathrm d\omega
  \end{split}.
\end{equation}

\begin{proof}
Considering that $f(x)=\mathrm{log}_2\left(1+\frac{1}{x}\right)$ is easily proofed to be a convex function, based on the Jensen's inequality the average achievable rate of $\mathrm{UE}_k$ is derived as
\begin{align}\label{er1}
\nonumber  R_{ik\_{\rm avg}} &= {\mathbb E}_{\omega_{iki}}\left[\mathbb E_{I_{ik}}\left[\mathrm{log}_2\left(1+\frac{M^2P_fP_p\omega_{iki}^2}{\left|x_{\mathrm{int}}
\right|^{2\alpha}
\left(I_{ik}+\sigma_n^2\right)}\right)\right]\right] \\
\nonumber  &\geq \mathbb E_{\omega_{iki}}\left[\mathrm{log}_2\left(1+\frac{M^2P_fP_p\omega_{iki}^2}{\left|x_{\mathrm{int}}
\right|^{2\alpha}
\mathbb E\left[\left(I_{ik}+\sigma_n^2\right)\right]}\right)\right]\\
&= \mathbb E_{\omega_{iki}}\left[\mathrm{log}_2\left(1+\frac{M^2P_fP_p\omega_{iki}^2}{\left|x_{\mathrm{int}}
\right|^{2\alpha}\left(I_{k\_{\rm avg}}+\sigma_n^2\right)}\right)\right].
\end{align}

Assumed that $\omega_{iki}$ is the shadowing effect in wireless channels and is calculated by $\omega_{iki}=e^{\frac{s_{iki}}{10}}$ with the random variable $s_{iki}$ governed by the Gaussian distribution with zero mean and variance $\sigma_s^2$, {\em i.e.}, $s_{iki}\sim N(0, \sigma_s^2)$, the probability density function (PDF) of $\omega_{iki}$ is derived as
\begin{equation}\label{wpdf}
  f_{\omega_{iki}}\left(\omega\right)= \frac{5}{\omega\sqrt{2\pi}\sigma_s}\mathrm {exp}\left(-\frac{25\ln^2(\omega)}{2\sigma_s^2}\right).
\end{equation}
Then the first moment (or the expectation) of $\omega_{iki}$ is given as
\begin{equation}\label{ew1}
  \mathbb E\left[\omega_{iki}\right]=e^{\frac{\sigma_s^2}{2}}.
\end{equation}
And the second moment of $\omega_{iki}$ is given as
\begin{equation}\label{ew2}
  \mathbb E\left[\omega_{iki}^2\right]=e^{2\sigma_s^2}.
\end{equation}
Assumed that the shadowing effects $\omega_{iki}$ and $\omega_{uki}(u\neq i)$ are {\em i.i.d.} random variables, we have $\mathbb E\left[\omega_{uki}^2\right]=\mathbb E\left[\omega_{iki}^2\right]=e^{2\sigma_s^2}$ and $\mathbb E\left[\omega_{uki}\right]=\mathbb E\left[\omega_{iki}\right]=e^{\frac{\sigma_s^2}{2}}$.

Substituting (\ref{eI}) and (\ref{wpdf}) into (\ref{er1}), the lower bound of the average achievable rate of $\mathrm{UE}_k$ in the $i$--th small cell can be obtained.
\end{proof}
\end{theorem}

In this paper, an UE is assumed to be associated with the closest BS and the distance vector between the UE and the associated BS is denoted as ${x_{\mathrm{int}}}$. Based on the result in~\cite{Alfano14J}, the PDF of the distance $\left|x_{\mathrm{int}}\right|$ is approximated by
\begin{equation}\label{fr}
  f_{x_{\mathrm{int}}}\left(r\right) \approx 2r\frac{1-e^{-\lambda_B\pi\delta^2}}{\delta^2}e^{-\lambda^{\ast}M(r,\delta)},
\end{equation}
with
\begin{equation}\label{M}
\begin{split}
&M(r,\delta)=\\
&\left\{\!
                \begin{array}{ll}
                  \pi r^2\!\!-\!\!\left(\!2\arcsin\left(\!\frac{\delta}{2r}\right)\!\!+\!\!\arccos\left(\frac{\delta}{2r}\right)\right)
                  r^2\!\!+\!\!\delta\sqrt{r^2\!\!-\!\!\frac{\delta^2}{4}},\;\hbox{$r\!>\!\frac{\delta}{2}$} \\
                  0,\;\hbox{$0\!<r\!\leq\frac{\delta}{2}$}
                \end{array}
              \right.
\end{split},
\end{equation}
where $\lambda^{\ast}$ is the density of the active BSs after adopting the BS switch-off strategy, and the value of $\lambda^{\ast}$ is the only one that guarantees $f_{x_{\mathrm{int}}}\left(r\right)$ to be a proper PDF function, {\em i.e.}, it is the unique value $\lambda^{\ast}$ such that $\int_0^\infty f_{x_{\mathrm{int}}}\left(r\right)\mathrm dr = 1$.

Moreover, the average sum achievable rate of the $i$--th cell in the thinning small cell network is calculated as
\begin{equation}\label{sr}
  \mathbb{E}(R_i)\!= \!\sum\limits_{k = 1}^K\mathbb{E}_{x_{{\mathrm {int}}}}\left[R_{ik\_{\rm avg}}\left(x_{{\mathrm {int}}}\right)\right].
\end{equation}
Considering that the $K$ UEs in the $i$--th cell are independently uniformly distributed in the coverage area of $\mathrm{BS}_i$, the average achievable rates of the $K$ UEs are equal, such that the average sum achievable rate of the $i$--th cell is further derived as
\begin{equation}\label{sr2}
  \mathbb{E}(R_i)\!=\!K\int\limits_0^\infty R_{ik\_{\rm avg}}\left(r\right)f_{x_{\mathrm{int}}}\left(r\right)\mathrm dr.
\end{equation}

Based on the signal~(\ref{xi}) transmitted by the BS with the massive MIMO technology, the transmission power of $\mathrm{BS}_i$ is
\begin{equation}\label{Pi}
  \begin{aligned}
    P_i & = \mathbf{x}_{i}^{\dag}\mathbf{x}_{i}\\
     & = \left[\sqrt{P_p}\left(\mathbf{H}_{ii}+\sum_{u^{'}\neq i}\mathbf{H}_{iu^{'}}\right)^{\ast}\mathbf{s}_{i}\right]^\dag\\
     &\times\sqrt{P_p}\left(
\mathbf{H}_{ii}+\sum_{u^{'}\neq i}\mathbf{H}_{iu^{'}}\right)^{\ast}\mathbf{s}_{i}\\
& = P_p\mathbf{s}_{i}^\dag\left(\sum\limits_{x_u}\mathbf{H}_{iu}^\dag
\mathbf{H}_{iu}\right)\mathbf{s}_{i}\\
& = MP_p\sum\limits_{x_u\in\Pi_{\mathrm{HCPP}}}\sum\limits_{k=1}^K\beta_{iku}
  \end{aligned}.
\end{equation}
Considering the $K$ UEs in the $i$--th cell are independently uniformly distributed in the coverage area of $\mathrm{BS}_i$, the distances between $\mathrm{BS}_i$ and $K$ UEs are {\em i.i.d.} random variables, and the PDF of the distances could be modeled by~(\ref{fr}).
The average BS transmission power of the small cell network is given as
\begin{equation}\label{Pavg}
  \begin{split}
    &P_{T\_{\rm avg}}  = \mathbb{E}\left[MP_p\sum\limits_{x_u}\sum\limits_{k=1}^K\beta_{iku}\right] \\
& = MP_p\sum\limits_{k=1}^K\mathbb{E}\left[\sum\limits_{x_u}\frac{\omega_{iku}}
{\left|x_{{\mathrm {BS}}_u}-x_{{\mathrm {BS}}_i}-x_{{\mathrm {int}}}\right|^{\alpha}}\right]\\
& = MP_p\sum\limits_{k=1}^K\mathbb{E}\left[\omega_{iku}\right]\mathbb{E}_{x_{\mathrm {int}}}\left[\int\limits_{\mathbb{R}^2}
\left|x+x_{{\mathrm {int}}}\right|^{-\alpha}\zeta ^{(2)}(x)\mathrm dx\right]
  \end{split}.
\end{equation}

Based on the decomposition of the total BS power consumption~\cite{Ge15J}, a linear average BS power consumption model is simple presented as follows
\begin{equation}\label{P}
  P_{\mathrm {BS}\_{\rm avg}}= \frac{P_{T\_{\rm avg}}}{\eta}+MP_{\mathrm{RF\_chain}}+P_{\mathrm{sta}},
\end{equation}
where $\eta $ is the average efficiency of signal transmission circuits and $P_{\mathrm{RF\_chain}}$ is the power of radio frequency circuit consumed at an antenna, $P_{\mathrm{sta}}$ is the BS operation power fixed as a constant. Furthermore, the average energy efficiency of the small cell network adopting the BS switch-off strategy is derived as (\ref{ee}) shown at the top of the next page.
\begin{figure*}[!t]
\begin{equation}\label{ee}
\mathrm {EE}= \frac{\lambda^{\ast}\mathbb{E}(R_i)}{\lambda^{\ast}P_{\mathrm {BS}\_\mathrm{avg}}}= \frac{L\lambda_B\int\limits_0^\infty R_{ik\_{\rm avg}}\left(r\right)f_{x_{\mathrm{int}}}\left(r\right)\mathrm dr}{\lambda^{\ast}\left(\frac{MP_p\sum\limits_{k=1}^Ke^{\frac{\sigma_s^2}{2}}\mathbb{E}_{x_{\mathrm {int}}}\left[\int\limits_{\mathbb{R}^2}
\left|x+x_{{\mathrm {int}}}\right|^{-\alpha}\zeta ^{(2)}(x)\mathrm dx\right]}{\eta}+MP_{\mathrm{RF\_chain}}+P_{\mathrm{sta}}\right)}
\end{equation}
\noindent\rule{\textwidth}{.5pt}%\vskip3pt
\end{figure*}

\section{Coverage Efficiency}\label{sec5}
The energy efficiency of 5G small cell networks can be optimized by effective BS switch-off approaches to saving energy in 5G small cell networks. However, the loss of coverage efficiency is an inevitable cost for 5G small cell networks when a part of small cell BSs are switched off. For the telecommunication providers, the coverage efficiency of small cell networks must keep a high level considering the user¡¯s experiences. Therefore, it is an important challenge to trade off the energy and coverage efficiency in 5G small cell networks. In this section, the coverage efficiency is analyzed to insight into the trade-off the energy and coverage efficiency of small cell networks.

According to the measurements in \cite{Meier91C}, the packet arrival process of a single user exhibited the long-range dependence characteristic. Through traffic analysis, Mah et al. \cite{Mah97C} determined statistics and distributions for the size of Hypertext Transfer Protocol (HTTP) files and the results showed that HTTP reply sizes have a heavy-tailed distribution. Furthermore, Park et al. \cite{Park96R} showed that in a "realistic" client/server network environment, {\em i.e.}, one with limited resources leading to the coupling of multiple traffic sources contending for shared resources, the degree to which file sizes are heavy-tailed directly determines the degree of traffic self-similarity. Therefore, the traffic of a single user has the self-similar and long-range dependent property in wireless networks. What's more, several mathematical distributions with the infinite variance have been proposed to fit the self-similar traffic, in which Pareto distributions with infinite variance \cite{Silva02C} have been widely used due to the analytical expression and intuitionistic engineering implication. In this case, the traffic at a UE is assumed to be governed by Pareto distribution in this paper. Moreover, the PDF of traffic at the UE is expressed as
\begin{equation}\label{traffic}
  f_{\rho}(\chi)=\frac{\theta\rho_{\min}^\theta}{\chi^{\theta+1}},\quad \chi \geq \rho_{\min},
\end{equation}
where $\theta\in(1,2]$ denotes the heavy tail coefficient of Pareto distribution, and $\rho_{\min}$ denotes the minimum transmission rate which satisfies the UE traffic requirement. Furthermore, the average traffic at the UE is given by
\begin{equation}\label{etraffic}
  \mathbb E\left[\rho\right]=\frac{\theta\rho_{\min}}{\theta-1}.
\end{equation}

When the BS switch-off strategy is adopted, the distance between a UE and the associated BS is larger, such that the transmission rate of the UE in the thinning small cell is reduced. Without loss of generality, a typical UE $\mathrm{UE}_0$ is considered at the origin of coordinates. In this paper, the coverage efficiency is defined as the probability that the transmission rate of the typical UE is larger than the UE's traffic requirement, which is expressed as
\begin{equation}\label{CE}
  \begin{aligned}
  \mathrm {CE} &= \mathrm{Pr}\left\{R_0>\rho_0\right\}\\
     &= \mathrm{Pr}\left\{\log_2\left(1+\mathrm{SINR}_0\right)>\rho_0\right\}\\
     &= \mathrm{Pr}\left\{\mathrm{SINR}_0>2^{\rho_0}-1\right\}
  \end{aligned},
\end{equation}
where $\rho_0$ is the traffic at the $\mathrm{UE}_0$, and $\mathrm{SINR}_0$ is the signal to interference plus noise ratio (SINR) at the $\mathrm{UE}_0$.

\begin{theorem}\label{T2}
The PDF of the SINR at the $\mathrm{UE}_0$ in the small cell network can be given as
\begin{equation}\label{sinrpdf}
f_{\mathrm{SINR}}(\gamma)\!=\! 2g^{-1}\!\left(\gamma\right)\!\frac{1\!-\!e^{-\lambda_B\pi\delta^2}}{\delta^2}\!e^{-\lambda^{\ast}M(g^{-1}
\left(\gamma\right),\delta)}\!\left|\frac{\mathrm dg^{-1}\left(\gamma\right)}{\mathrm d\gamma}\right|.
\end{equation}

\begin{proof}
Based on (\ref{yik}), the received signal of the $\mathrm{UE}_0$ is expressed as
\begin{equation}\label{y0}
  y_0 = M\sqrt{P_fP_p}\beta_{00}s_0+M\sqrt{P_fP_p}\sum\limits_{u \neq 0}\beta_{u0}s_u+n_0,
\end{equation}
where $s_0$ and $s_u$ respectively denotes the transmission signal for the $\mathrm{UE}_0$ and the interference signal, $\beta_{00}$ and $\beta_{u0}$ denotes the large scale fading coefficient between BSs and the $\mathrm{UE}_0$, $n_0$ is the noise at the $\mathrm{UE}_0$. Furthermore, the $\mathrm{SINR}_0$ can be derived as
\begin{equation}\label{sinr1}
  \mathrm {SINR}_0 = \frac{M^2 P_fP_p\beta_{00}^2}{M^2P_fP_p\sum\limits_{u \neq 0}\beta_{u0}^2+\sigma_n^2}=\frac{M^2 P_fP_p\beta_{00}^2}{I_0+\sigma_n^2}.
\end{equation}
To simplify the complex calculation, the instantaneous interference $I_0$ at the $\mathrm{UE}_0$ is assumed to be approximatively equal to the average interference $I_{0\_\mathrm{avg}}$, which the average interference is given based on (\ref{eI}) as
\begin{equation}\label{eI0}
  I_{0\_\mathrm{avg}}(x_{\mathrm{int}})= \frac{M^{2}P_fP_pe^{2\sigma_s^2}}{\zeta^{(1)}}\int_{\mathbb R^2}\frac{\zeta^{(2)}\left(\left|x\right|\right)}{\left|x+x_{\mathrm {int}}\right|^{2\alpha}}\mathrm d{x},
\end{equation}
and
\begin{equation}\label{sinr2}
 \mathrm{SINR}_0 = \frac{M^2 P_fP_p\omega_{00}^2}{\left| x_{\mathrm {int}}\right|^{2\alpha}I_{0\_\mathrm{avg}}(x_{\mathrm{int}})+\left| x_{\mathrm {int}}\right|^{2\alpha}\sigma_n^2}.
\end{equation}
In this case, the SINR of the $\mathrm{UE}_0$ can be denoted as a function of $\left| x_{\mathrm {int}}\right|$, {\em i.e.}, $\mathrm{SINR}_0=g\left( \left|x_{\mathrm {int}}\right|\right)$. Moreover, the inverse function of $\mathrm{SINR}_0\left(\left|x_{\mathrm {int}}\right|\right)$ is denoted as $\left| x_{\mathrm {int}}\right|=g^{-1}\left(\mathrm{SINR}_0\right)$. Let $\gamma=\mathrm{SINR}_0$, the PDF of $\mathrm{SINR}_0$ can be derived as (\ref{sinrpdf}).
\end{proof}
\end{theorem}

Based on (\ref{CE}) and (\ref{traffic}), the coverage efficiency of the small cell network is expressed as
\begin{equation}\label{CE2}
  \mathrm{CE} = \int\limits_{2^{\rho}-1}^\infty f_{x_{\mathrm{int}}}\left(g^{-1}\left(\gamma\right)\right)\left|\frac{\mathrm dg^{-1}\left(\gamma\right)}{\mathrm d\gamma}\right|\mathrm d\gamma.
\end{equation}
When (\ref{fr}) is substituted into (\ref{CE2}), the coverage efficiency of the small cell network is extended as
\begin{equation}\label{CE3}
    \mathrm{CE}\!=\!\!\int\limits_{2^{\rho}\!-1}^\infty \!\! 2g^{-1}\!\left(\gamma\right)\frac{1\!\!-\!\!e^{-\!\lambda_B\pi\delta^2}}{\delta^2}e^{-\!\lambda^{\ast}M(g^{-1}
\left(\gamma\right),\delta)}\!\left|\!\frac{\mathrm dg^{-1}\left(\gamma\right)}{\mathrm d\gamma}\!\right|\mathrm d\gamma.
\end{equation}

\section{Simulation Results and Discussions}\label{sec6}

In this section, numerical results of the energy efficiency and coverage efficiency are analyzed for the small cell network. The impact of the BS density $\lambda_B$, the minimum distance between active BSs $\delta$, and the number of antennas at a BS $M$ on the energy and coverage performance of the thinning small cell network are simulated for performance analysis. The configured default parameters are in Table~\ref{tab1}.
\begin{table}[!b]
\processtable{Default Parameters\label{tab1}}
{\begin{tabular*}{20pc}{@{\extracolsep{\fill}}ll@{}}\toprule
Parameter  &Value  \\
\midrule
BS density $\lambda_B$  &$10^{-4}$  \\
Minimum distance between active BSs $\delta$  &200m  \\
Number of antennas at a BS $M$  &128   \\
Number of UEs in a small cell $L$ &5  \\
Shadowing standard deviation $\sigma_s$ &6dB  \\
Noise power $\sigma_n$ &-174dBm  \\
Path loss exponent $\alpha$ & 4 \\
Signal transmission power $P_f$ & 7.7W \\
Pilot signal power $P_p$ & 0.13W \\
Power amplification coefficient $\eta$ & 0.38 \\
Power of radio frequency circuit per antenna$P_{\mathrm{RF\_chain}}$ & 0.048W \\
BS operation power $P_{\mathrm{sta}}$ & 4.3W\\
\botrule
\end{tabular*}}{}
\end{table}

\begin{figure}
  \centering\includegraphics[width=8cm]{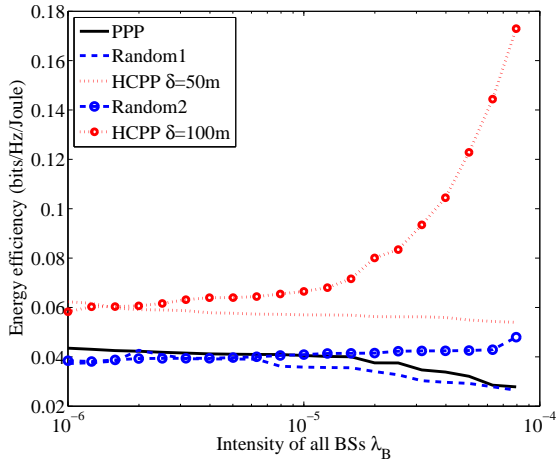}\\
  \caption{of the small cell network considering there different cases, {\em i.e.} "PPP" in the figure denotes the case that no BS is switched in the small cell network, "HCPP" denotes the case that the proposed BS switch-off strategy with mutual repulsion is adopted, "Random" denotes the case that the random switch-off strategy~\cite{Soh13C} is adopted with the switch-off probability $\frac{\lambda^{\ast}}{\lambda_B}$.}\label{f2}
\end{figure}

Fig.~\ref{f2} shows that the energy efficiency of the small cell network considering three different cases, {\em i.e.} "PPP" in the figure denotes the case that no BS is switched in the small cell network, "HCPP" denotes the case that the proposed BS switch-off strategy with mutual repulsion is adopted, "Random" denotes the case that the random switch-off strategy~\cite{Soh13C} is adopted with the switch-off probability $\frac{\lambda^{\ast}}{\lambda_B}$. The energy efficiency of the proposed strategy is larger than that of other two cases. The energy efficiency is improved more with the increase of the minimum distance compared with the two strategies. The energy efficiency increases with the increase of the intensity of all BSs when the minimum distance is fixed at 100m.

\begin{figure}
  \centering\includegraphics[width=8cm]{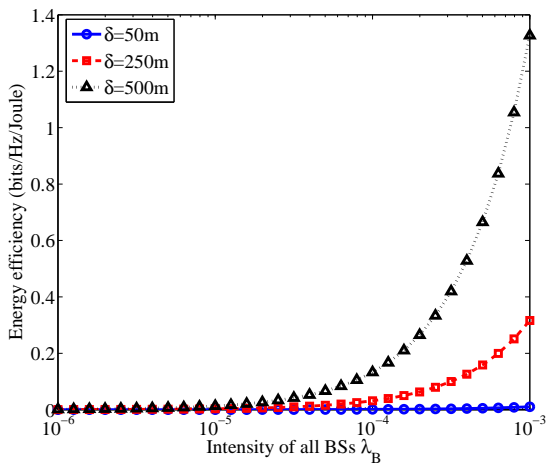}\\
  \caption{The energy efficiency of the thinning small cell network as functions of the intensity of all BSs, considering different minimum distances between active BSs.}\label{f3}
\end{figure}

Fig.~\ref{f3} shows that the energy efficiency of the thinning small cell network as functions of the intensity of all BSs, considering different minimum distances between active BSs. The energy efficiency increases with the increase of the intensity of all BSs in the small cell network. The minimum distance $\delta$ is larger, the energy efficiency increases more rapidly.

\begin{figure}
  \centering\includegraphics[width=8cm]{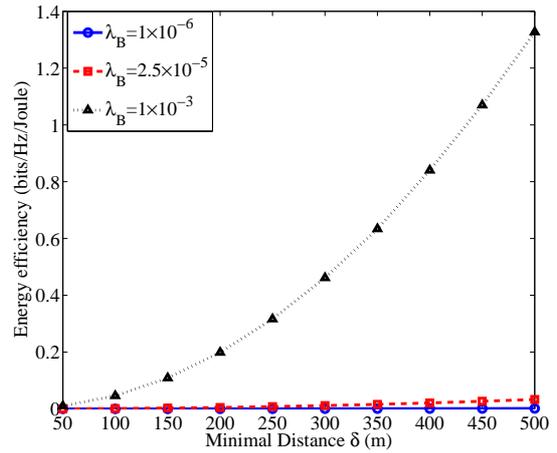}\\
  \caption{The energy efficiency of the thinning small cell network as functions of the minimum distance, considering different intensities of all BSs.}\label{f4}
\end{figure}

Fig.~\ref{f4} investigates that the energy efficiency of the thinning small cell network as functions of the minimum distance, considering different intensities of all BSs. The energy efficiency increases with the increase of the minimum distance $\delta$. The intensity $\lambda_B$ is larger, the impact of the BS switch-off strategy on improving the energy efficiency is larger. The energy efficiency of the small cell network can be improved by increasing the intensity of all BSs and increasing the minimum distance of the BS switch-off strategy with mutual repulsion. The energy efficiency increases more when larger intensity and larger minimum distance are adopted.

\begin{figure}
  \centering\includegraphics[width=8cm]{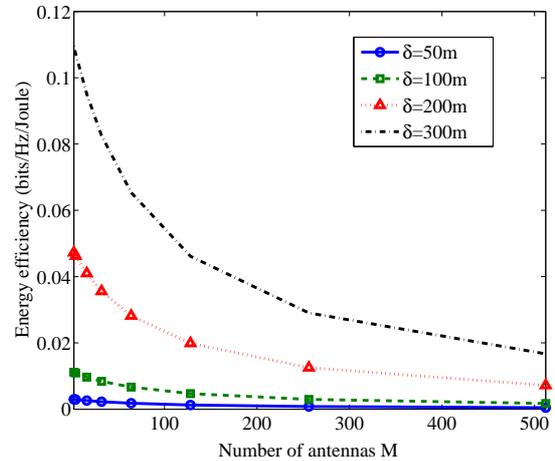}\\
  \caption{The energy efficiency of the thinning small cell network as functions of the number of antennas, considering different minimum distances.}\label{f5}
\end{figure}

Fig.~\ref{f5} illustrates that the energy efficiency of the thinning small cell network as functions of the number of antennas, considering different minimum distances. The energy efficiency decreases with the increase of the number of antennas $M$. The energy efficiency gains with the increase of the minimum distance are decreased with the increase of the number of antennas.

\begin{figure}
  \centering\includegraphics[width=8cm]{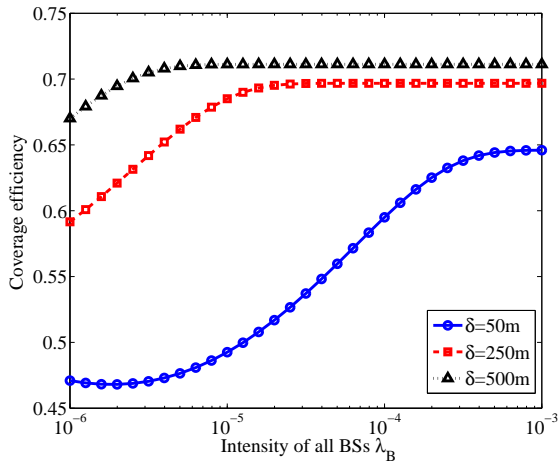}\\
  \caption{The coverage efficiency of the thinning small cell network as functions of the intensity of all BSs considering different minimum distances between active BSs.}\label{f7}
\end{figure}

Fig.~\ref{f7} shows that the coverage efficiency of the thinning small cell network as functions of the intensity of all BSs, considering different minimum distances between active BSs. When the minimum distance is fixed, {\em e.g.} $\delta=250m$, the coverage efficiency increases with the increase of the intensity when the intensity is smaller than $2.5\times10^{-5}$; the coverage efficiency keeps unchanged with the increase of the intensity when the intensity is larger than $2.5\times10^{-5}$. The value of the intensity at which the coverage efficiency achieves the maximum decreases with the increase of the minimum distance between active BSs.

\begin{figure}
  \centering\includegraphics[width=8cm]{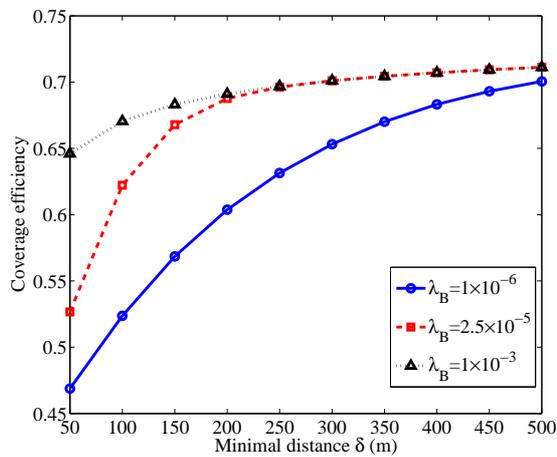}\\
  \caption{The coverage efficiency of the thinning small cell network as functions of the minimum distance between active BSs considering different intensities of all BSs.}\label{f8}
\end{figure}

Fig.~\ref{f8} illustrates that the coverage efficiency as functions of the minimum distance between active BSs, considering different intensities of all BSs. The coverage efficiency increases with the increase of the minimum distance. With the increase of the minimum distance, values of the coverage efficiency of different intensities are equal, {\em e.g.} the coverage efficiencies of the intensities $2.5\times10^{-5}$ and $1\times10^{-3}$ are equal when the minimum distance is larger than 250m. It is illustrated that the coverage efficiency is only limited by the minimum distance when the minimum distance is larger than 250m.

\begin{figure}
  \centering\includegraphics[width=8cm]{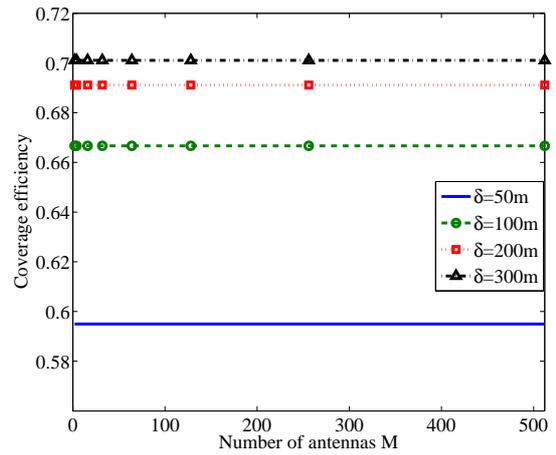}\\
  \caption{The coverage efficiency of the thinning small cell network considering different numbers of antennas and minimum distances.}\label{f9}
\end{figure}

Fig.~\ref{f9} shows that the coverage efficiency of the thinning small cell network considering different numbers of antennas and minimum distances. The coverage efficiency keeps unchanged with the increase of the number of antennas $M$. According to the expression of SINR as (\ref{sinr1}), it is found that the power gain effects due to the number of antennas $M$ on the desired signal power (the numerator) and the interfering signal power (the denominator) are equivalent, which results in that the SINR is independent of the number of antennas. Thus, the coverage efficiency is independent of $M$.

\section{Conclusion}\label{sec10}

Considering the BSs switch-off strategy, the energy and coverage efficiency of the thinning small cell network are analyzed in the massive MIMO system in this paper. Based on our results, the energy efficiency increases rapidly with the increase of the intensity of all BSs and the minimum distance between active BSs. And the maximum coverage efficiency can be achieved with the increase of the intensity and minimum distance. Moreover, the energy efficiency and coverage efficiency performance are better than the random strategy. The optimal coverage efficiency can be achieved when the minimum distance is 250m and the intensity is $2.5\times10^{-5}$. For the future study, the requirement of user QoS could be considered to further optimize the energy and coverage efficiency of 5G small cell networks.

\end{document}